\documentclass[twocolumn]{aastex63}

\def\feh{$\mathrm{[Fe/H]}$}

\usepackage{amsmath}

\hypersetup{linkcolor=red,citecolor=blue,filecolor=green,urlcolor=magenta}

\begin{document}

\shorttitle{Aquarius III \& RR Lyrae}
\shortauthors{Ngeow \& Bhardwaj}

\title{Candidate RR Lyrae Associated with the Ultrafaint Dwarf Galaxy Aquarius III}

\correspondingauthor{Chow-Choong Ngeow}
\email{cngeow@astro.ncu.edu.tw}

\author[0000-0001-8771-7554]{Chow-Choong Ngeow}
\affil{Graduate Institute of Astronomy, National Central University, 300 Jhongda Road, 32001 Jhongli, Taiwan}

\author[0000-0001-6147-3360]{Anupam Bhardwaj}
\affil{Inter-University Center for Astronomy and Astrophysics (IUCAA), Post Bag 4, Ganeshkhind, Pune 411 007, India}

\begin{abstract}

  We report the search of RR Lyrae in the vicinity of a newly discovered ultrafaint dwarf galaxy, Aquarius III. Based on the known RR Lyrae catalogs and $gri$-band light curves retrieved from public archives, we identified a RR Lyrae with distance, metallicity, and proper motion consistent with Aquarius III. Therefore, this RR Lyrae is the first variable star identified to be associated with Aquarius III, despite its projected distance is more than 15 times the half-light radius of Aquarius III. On the other hand, a dedicated time-series monitoring of the central part of Aquarius III, out to a projected radius of approximately four half-light radius, revealed there is no RR Lyrae in this region. We ran a set of synthetic color-magnitude diagrams with properties similar to Aquarius III, and found a non-negligible probability that Aquarius III could have (at least one) RR Lyrae. We have also identified a RR Lyrae candidate but most likely it is a background halo star.

\end{abstract}


\section{Introduction}\label{sec1}

It is of great interest to discover pulsating stars, especially RR Lyrae, in the dwarf galaxies and the ultrafaint dwarf galaxies \citep[UFD; for a review of UFD, see][]{simon2019}, because they serve as distance, metallicity, and age indicators, and constrain properties of old stellar populations in their host galaxies. For recent reviews on the importance of pulsating stars for the studies of dwarf and UFD galaxies, see \citet{monelli2022} and \citet{mv2023}. 

A new UFD, Aquarius III, was discovered by \citet{cerny2024} using the DECam Local Volume Exploration \citep[DELVE,][]{dw2021} survey data. Some of the properties of this UFD, summarized in Table \ref{tab1}, suggest that it is remarkably similar to Virgo III \citep{homma2024} in terms of their stellar mass, half-light radius ($r_h$), and absolute $V$-band magnitude ($M_V$). Based on the empirical relation between the expected number of RR Lyrae and $M_V$ presented in \citet{mv2023}, both Virgo III and Aquarius III are expected to have $1\pm1$ RR Lyrae. Indeed, \citet{ngeow2024} discovered not one, but three RR Lyrae associated with Virgo III. Therefore, detection or non-detection of RR Lyrae in Aquarius III would be useful for the study of such old pulsating stars in the UFD, especially those with $M_V \gtrsim -5.0$~mag.

In Section \ref{secrrl}, we present our search results of RR Lyrae in Aquarius III. We further evaluate the detectability of RR Lyrae in UFD with properties similar to Aquarius III and Virgo III, using a set of synthetic color-magnitude diagrams (CMD), in Section \ref{seccmd}, followed by the conclusion of this work in Section \ref{seclast}.   

\begin{deluxetable}{lccc}
  \tabletypesize{\scriptsize}
  \tablecaption{Comparison of Virgo III and Aquarius III.\label{tab1}}
  \tablewidth{0pt}
  \tablehead{
    \colhead{Property} &
    \colhead{Unit} &
    \colhead{Virgo III} &
    \colhead{Aquarius III}
    }
  \startdata
  Reference       & $-$ & \citet{homma2024} & \citet{cerny2024} \\
  $D$             & kpc     & $151^{+15}_{-13}$ & $85\pm4$ \\
  $\epsilon$      & $-$ & $0.29^{+0.15}_{-0.19}$ & $0.47^{+0.14}_{-0.28}$ \\
  $M_V$           &  mag    & $-2.7^{+0.5}_{-0.6}$ & $-2.5^{+0.3}_{-0.5}$ \\
  $r_h$           &  pc     & $44^{+14}_{-12}$    & $41^{+9}_{-8}$  \\
  $M_*$          &$M_\odot$ & $1500^{+700}_{-500}$ & $1700^{+760}_{-520}$ \\
  $N_{BHB}$        & $-$     & 3  & 1 \\
  $\tau$\tablenotemark{a}          & Gyr     & 13  & 13.5\\
  $\mathrm{[M/H]}$ or \feh         & dex     & $-2.2$\tablenotemark{a} & $-2.61\pm0.21$ \\
  \enddata
  \tablenotetext{a}{Assumed or fixed during the isochrone fitting to the observed CMD.}
  \tablecomments{$D=$ Heliocentric distance; $\epsilon=$ Ellipticity; $M_V=$ Absolute $V$-band magnitude; $r_h=$ Half-light radius;  $M_*=$ Stellar mass; $N_{BHB}=$ number of blue horizontal-branch (BHB) stars; $\tau=$ Age; $\mathrm{[M/H]}$ or \feh$=$ Metallicity.}
\end{deluxetable}

\section{A Search of RR Lyrae Variable in Aquarius III} \label{secrrl}

\subsection{Based on Photometric Monitoring with LOT} \label{seclot}

\begin{figure}
  \epsscale{1.1}
  \plotone{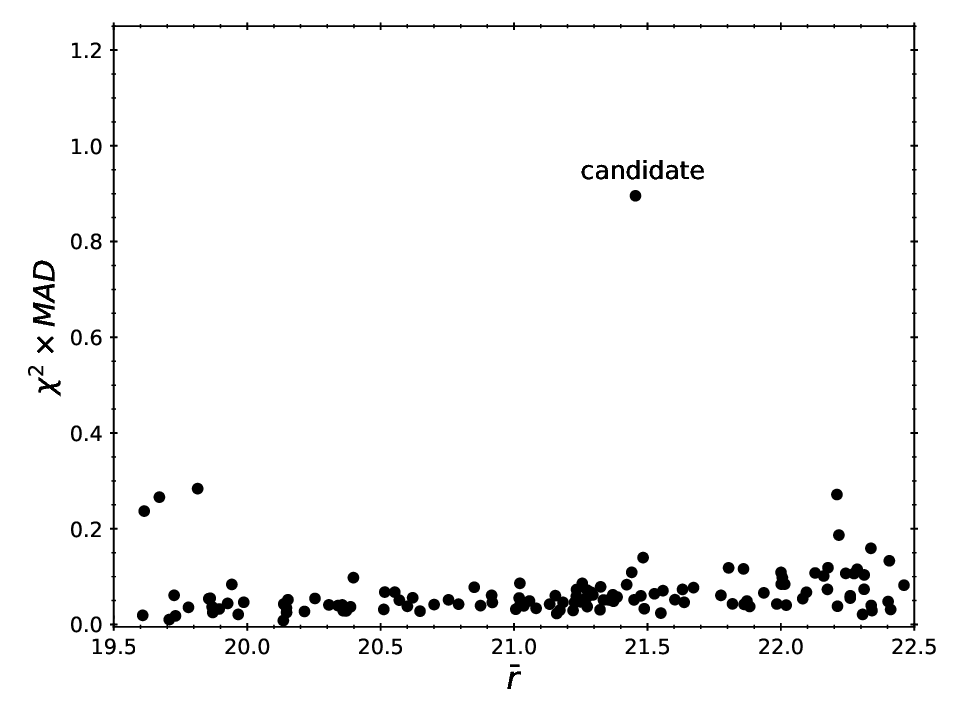}
  \caption{The values of $\chi^2\times MAD$ as a function of mean $r$-band magnitudes for the SDSS DR16 stars with more than 10 data-points on the LOT $r$-band light curves. A variable star candidate clearly stands out against the constant stars with $\chi^2\times MAD \sim 0.05$. Three bright stars with $\bar{r}\sim 19.6$~mag and $\chi^2\times MAD \sim 0.25$ appear to be low-amplitude variable stars. However, based on their light curves and locations on the CMD, they are clearly not RR Lyrae, nor the interest of this work.}
  \label{fig_cm}
\end{figure}

We carried out time-series observations, using the Lulin One-meter Telescope (LOT), on Aquarius III from 06 November to 07 December, 2024. In total, we collected 64 $r$-band images with 600~second exposure time. The image reduction and the photometric calibration of these LOT images were identical to \citet{ngeow2024}, interested readers are advised to refer to \citet{ngeow2024} for more details. The median seeing of these images is $1.26\arcsec$, reaching a depth of $r\sim 22.0$~mag. A major difference between this run of observations and in \citet{ngeow2024} was that our observations were carried out using the Princeton Instruments SOPHIA 2048B CCD camera, resulting a slightly larger field-of-view (FOV) of $13.1\arcmin \times 13.1\arcmin$, with a pixel scale of $0.385\arcsec$~pixel$^{-1}$. 

\begin{figure}
  \epsscale{1.1}
  \plotone{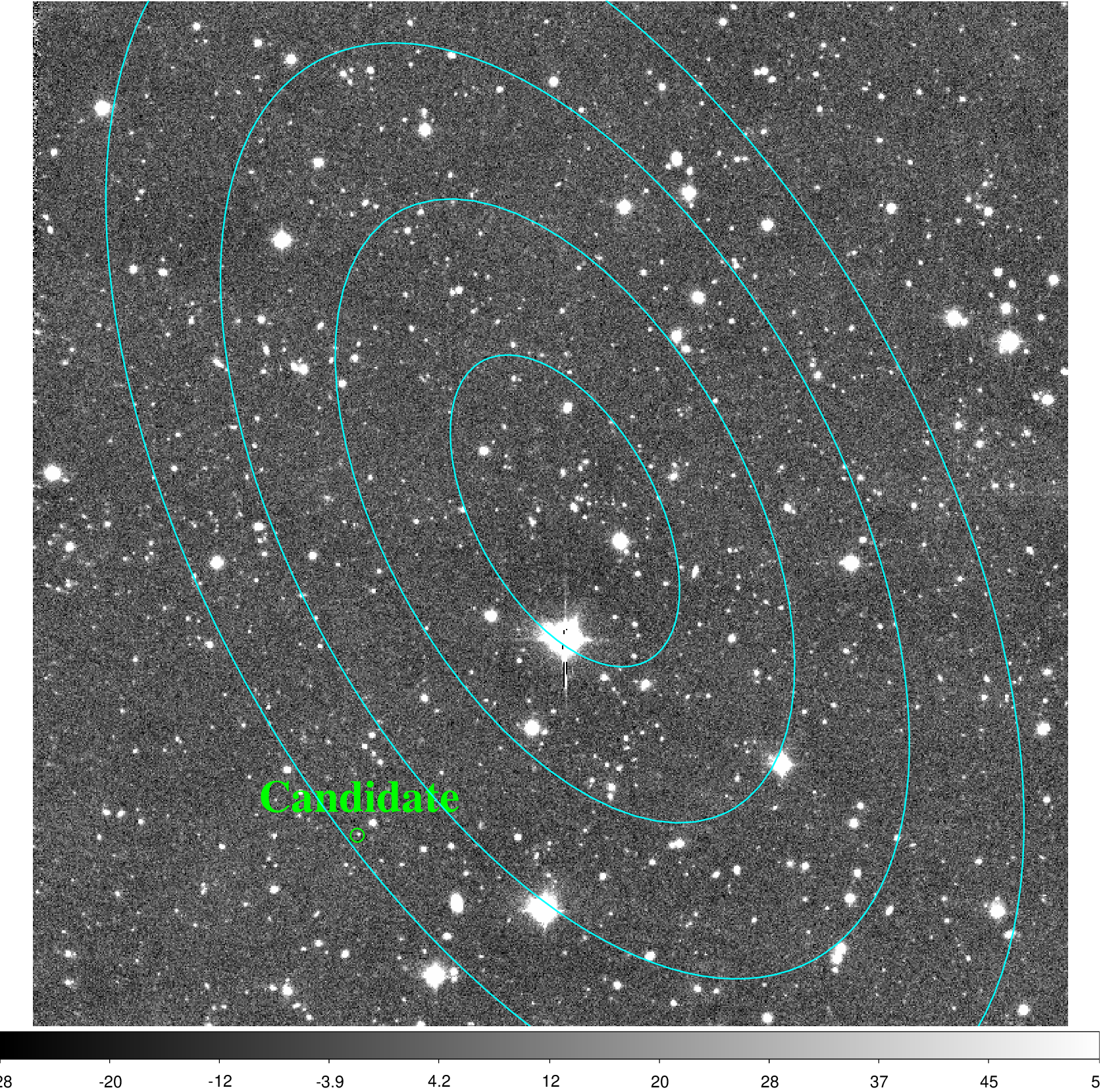}
  \caption{A coadded image by median-combined all of the LOT images using {\tt SWARP} \citep{bertin2002}. Location of the candidate variable star is also marked on the image. The cyan contours represent $\{1,\ 2,\ 3,\ 4\} \times r_h$ \citep[with parameters taken from][]{cerny2024}. }
  \label{fig_coadd}
\end{figure}

We cross-matched the detected sources in these LOT image with stars classified in the Sloan Digital Sky Survey (SDSS) Data Release 16 (DR16) catalog \citep{sdss2020}, and there are about $230$ stars located within the LOT images. After calibrating the instrumental point-spread function (PSF) magnitudes for these stars to the Pan-STARRS1 \cite[PS1,][]{chambers2016,flewelling2020} AB photometric system, we constructed their $r$-band light curves and searched for potential variable stars using the $\chi^2\times MAD$ metric \citep[for more details, see][]{ngeow2024}. From Figure \ref{fig_cm}, we identified a candidate variable star at $r\sim 21.5$~mag. Its location with respect to Aquarius III is shown in Figure \ref{fig_coadd}, which is $\sim 4r_h$ away from the center of Aquarius III. 

However, the brightness of this candidate variable star is $\gtrsim 1$~mag fainter than the expected brightness of RR Lyrae in Aquarius III (around $r\sim20.2$~mag). Nevertheless, by combining the LOT and archival light curves, we found that this candidate variable star is most likely a background RR Lyrae at a distance of $\sim 163$~kpc. Details on the analysis of this candidate RR Lyrae can be found in the Appendix. Our dedicated search suggested there is no RR Lyrae associated with Aquarius III within its $\sim 4r_h$ area (as covered by the LOT observations). On the other hand, \citet{tau2024} shows that RR Lyrae for some of the UFD could be located beyond $4r_h$ from the host galaxies. Hence, in the next subsection we extended our search to a larger area.

\subsection{Based on Archival Photometric Data}\label{secet}

Following the approach of \citet{tau2024}, we searched for known RR Lyrae within $16r_h$ radius of Aquarius III from the PS1 $3\pi$ RR Lyrae Catalog \citep{sesar2017} and the Gaia Data Release 3 \citep[DR3,][]{gaia2016,gaia2023} RR Lyrae Catalog \citep{clementini2023}. An ab-type (or fundamental mode) RR Lyrae was identified at $\sim25\arcmin$ away from Aquarius III, with a PS1 ID of 103753576326015348 or a Gaia DR3 ID of 2447369809479339136. 

From Gaia DR3 catalog, the measured proper motion for this RR Lyrae is $(\mu_{\alpha*},\ \mu_\delta) = (0.923\pm0.637,\ -0.549\pm0.565)\ \mathrm{mas} \ \mathrm{yr}^{-1}$. This proper motion is consistent to the proper motion of Aquarius III reported in \citet{cerny2024}, with $(\mu_{\alpha*},\ \mu_\delta)_{\mathrm{AquIII}} = (1.01\pm0.25,\ -0.10\pm0.20)\ \mathrm{mas} \ \mathrm{yr}^{-1}$. Furthermore, the extinction corrected mean $r$-band magnitude for this RR Lyrae, reported in the PS1 $3\pi$ RR Lyrae Catalog \citep{sesar2017}, is $20.05$~mag,\footnote{Note that all of the mean magnitudes in the PS1 $3\pi$ RR Lyrae Catalog have been corrected for the extinction, using the dust map from \citet{schlafly2014}; for more details, see \citet{sesar2017}. Specifically, the dust map returned a $E(B-V)=0.078$~mag for this RR Lyrae. The corresponding mean magnitude without the extinction correction is $20.23$~mag.} consistent with the expected brightness of a RR Lyrae at the distance of Aquarius III. Therefore, this RR Lyre could be a RR Lyrae of Aquarius III.

\begin{figure}
  \epsscale{1.2}
  \plotone{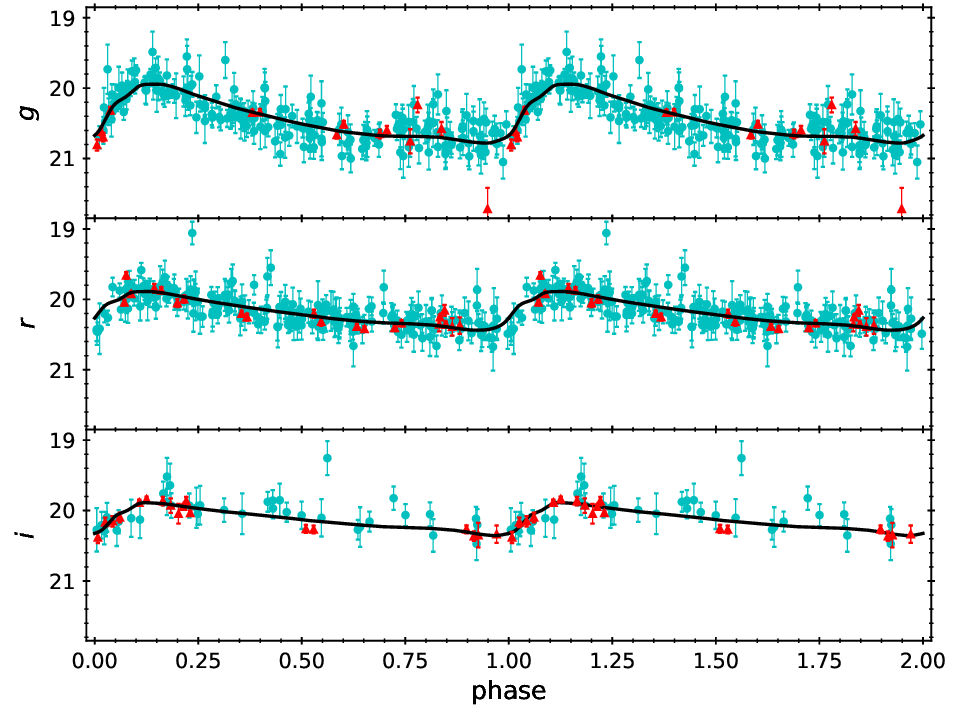}
  \caption{The $gri$-band phased light curves for the potential RR Lyrae identified in Section \ref{secet}. The cyan filled circles and red triangles are the data-points retrieved from ZTFDR22 and PS1DR2, respectively. The black curves are the best-fit template light curves derived in \citet{braga2024}.}
  \label{fig_lc}
\end{figure}

We retrieved the $gri$-band light curves for this RR Lyrae from the \dataset[PS1 Data Release 2]{http://dx.doi.org/10.17909/s0zg-jx37} \citep[PS1DR2,][]{stsci2022} archive\footnote{The API to access the PS1DR2 data is given at \url{https://ps1images.stsci.edu/ps1_dr2_api.html}} and the Zwicky Transient Facility \citep[ZTF,][]{bellm2019,graham2019} \dataset[Data Release 22]{http://dx.doi.org/10.17909/s0zg-jx37} (ZTFDR22). Note that the ZTF photometry has been calibrated to the PS1 system \citep{masci2019}, hence both sets of light curves are in the same AB photometric system. The numbers of data-points in the $g/r/i$-band retrieved from PS1DR2 and ZTFDR22 are 15/22/19 and 211/225/39, respectively. Using the multi-band periodogram subroutines from the {\tt gatspy} \citep{vdp2015} python package, we refined the pulsation period ($P$) of this RR Lyrae to be $P= 0.648708$~days, which is slightly shorter than the period reported in the PS1 $3\pi$ RR Lyrae Catalog ($P=0.648710$~days) or the Gaia DR3 RR Lyrae Catalog ($P=0.648712$~days). Figure \ref{fig_lc} presents the $gri$-band light curves folded with our refined period.

By fitting a set of template light curves derived in \citet{braga2024}\footnote{The fitting subroutines are available at \url{https://github.com/vfbraga/RRL_lcvtemplate_griz_LSST/tree/main}} to the observed $gri$-band light curves shown in Figure \ref{fig_lc}, the $gri$-band mean magnitudes are found to be $20.446\pm0.009$~mag, $20.192\pm0.007$~mag, and $20.134\pm0.014$~mag, respectively. We have also fit the $gr$-band light curve using the 8th-order Fourier expansion (in sine series) to derive the Fourier parameter $\phi_{31}(=\phi_3 - 3\phi_1)$, as $\phi^g_{31}=4.874\pm0.200$~rad and $\phi^r_{31}=4.891\pm0.275$~rad. Together with the $gr$-band \feh-$P$-$\phi_{31}$ relations derived in \citet{ngeow2022a}, the photometric metallicity for this RR Lyrae was found to be $-2.16\pm0.24$~dex and $-2.60\pm0.30$~dex, based on the $gr$-band light curves, respectively. Averaging these two metallicities we obtained $-2.38\pm0.39$~dex for this RR Lyrae, where the error is based on small number statistics \citep{keeping1962}. This photometric metallicity is consistent with the spectroscopic metallicity of Aquarius III \citep[$-2.61\pm0.21$~dex,][]{cerny2024}.  

With the period, photometric metallicity, and mean magnitudes obtained from the multi-band light curves, we derived the distance to this RR Lyrae using the period-Wesenheit-metallicity (PWZ) relations found in \citet{ngeow2022}. We prefer to use the PWZ relations because the Wesenheit magnitude is extinction-free by construction \citep{madore1991}. The derived distance moduli using the $gr$-band, the $ri$-band, and the $gi$-band PWZ relations are $19.60\pm0.21$~mag, $19.83\pm0.16$~mag, and $19.73\pm0.16$~mag, respectively. Averaging out these distance moduli we obtained $19.72\pm0.14$~mag (the error is based on the small number statistics), which translates to a linear distance of $87.9\pm5.7$~kpc.\footnote{Oddly, the parallax listed in the Gaia DR3 Catalog is $1.4323$~mas, a value that is too large at the distance of Aquarius III.} Again, this distance is fully consistent with the distance of Aquarius III \citep[$85\pm4$~kpc,][]{cerny2024}.

Since the distance, proper motion, and metallicity of this RR Lyrae are all consistent with Aquarius III, we concluded this RR Lyrae is the first RR Lyrae associated with Aquarius III. We propose renaming this RR Lyrae as Aquarius III-V1. The ultimate confirmation of its membership has to wait for the (multi-epoch) radial velocity measurements.

\section{The Synthetic CMD} \label{seccmd}

\begin{figure*}
  \epsscale{1.1}
  \plottwo{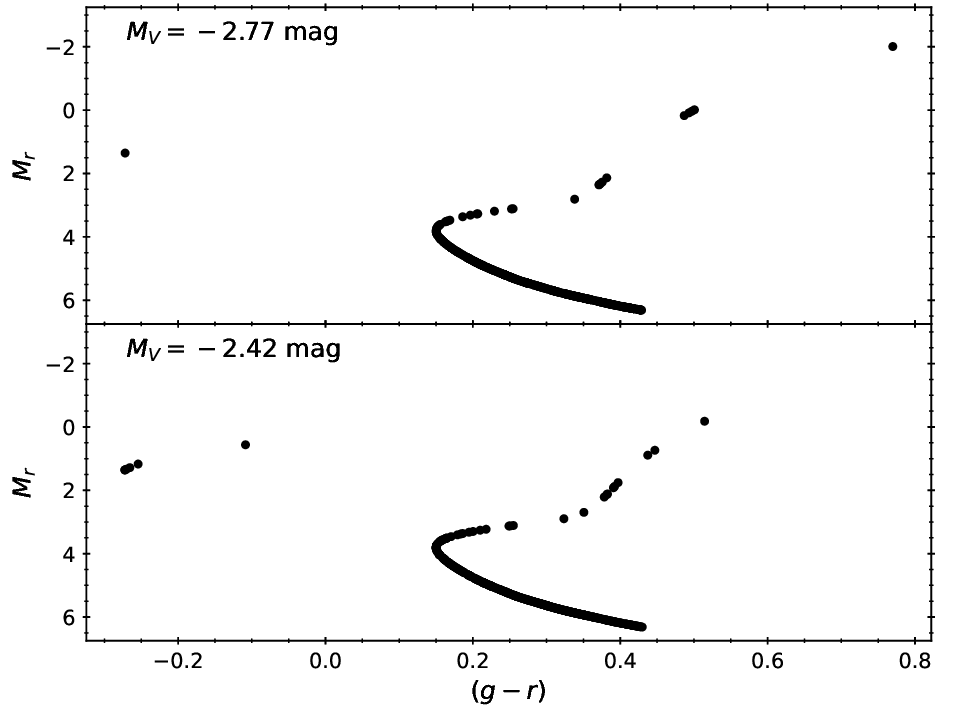}{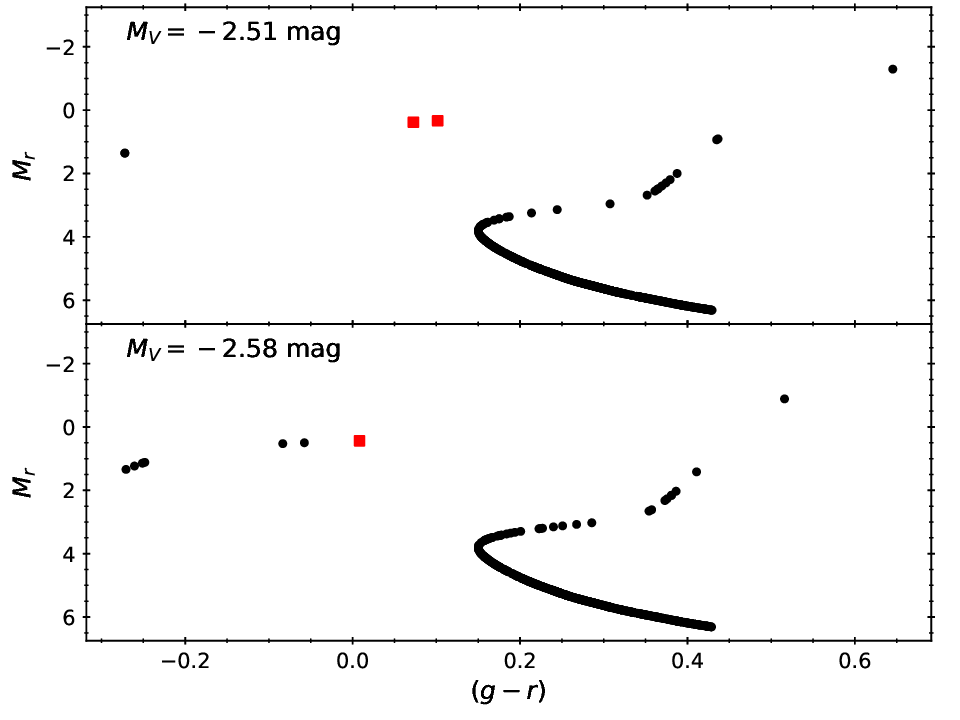}
  \caption{Two examples of the synthetic CMD without (left panels) and with (right panels, shown as red squares) RR Lyrae. These synthetic CMD were constructed using the {\tt BaSTI} web-tool, which also provide the identification of RR Lyrae from the synthetic stars. These four synthetic CMD have a slightly different input SFR scales and resulted a different $M_V$ (see text for more details).}
  \label{fig_scmd}
\end{figure*}

Using LOT, which has a relatively small FOV, we found three RR Lyrae in Virgo III \citep{ngeow2024} but none in Aquarius III (Section \ref{seclot}). As discussed in Section \ref{secet}, Aquarius III indeed has a RR Lyrae but located quite far from Aquarius III. On the other hand, \citet{ngeow2025} pointed out that for UFD with $M_V\gtrsim -5.0$~mag, the chance of a UFD to have at least one RR Lyrae is about half. This motivated us to investigate the detectability of RR Lyrae in UFD for those UFD with similar properties to Virgo III and Aquarius III. The {\tt BaSTI} \citep[a Bag of Stellar Tracks and Isochrones,][]{basti2021} suite of tools include a web interface\footnote{\url{http://basti-iac.oa-abruzzo.inaf.it/syncmd.html}} to generate synthetic CMD for a set of user-specific parameters. Furthermore, the output of {\tt BaSTI} synthetic CMD can include the number of pulsating stars, such as RR Lyrae and Cepheids, if there are synthetic stars falling within the theoretical instability strip \citep[for more details, see][]{basti2021}. 

Using this {\tt BaSTI} online tool, we generated 100 synthetic CMD with properties similar to Virgo III and Aquarius III (see Table \ref{tab1}). Specifically, we adopted the $\alpha$-enhanced models, helium abundance of $Y=0.247$, the normalize star formation rate (SFR) of 1, age of 13~Gyr, metallicity of $-2.4$~dex (the mid-point value of Virgo III and Aquarius III), and using the \citet{kroupa1993} initial mass function. For simplicity, we assume no metallicity spread and no binary fraction on these synthetic CMD. We set the low mass limit to be $0.6M_\odot$, because this mass is approaching the lower limit for RR Lyrae. The only parameter we altered is the SFR scale (the number of stars per age-bin, between 1300 and 1400) such that the resulted synthetic CMD has a total stellar mass of $\sim 2000M_\odot$, and its $M_V$ is between $\sim -2.8$~mag and $\sim -2.4$~mag. From these set of 100 synthetic CMD, there are 27 and 7 of them have one and two RR Lyrae, respectively. Examples of (randomly selected) two synthetic CMD with and without RR Lyrae are shown in Figure \ref{fig_scmd}. This suggested UFD with $M_V$ similar to Virgo III and Aquarius III has a non-negligible probability to host RR Lyrae.  

\section{Conclusion} \label{seclast}

The main goal of our work is to search for RR Lyrae associated with Aquarius III UFD. Using the LOT time-series observations, we identified a candidate RR Lyrae with projected angular distance of $\sim 4r_h$ away from Aquarius III. However, this candidate RR Lyrae turned out to be a background halo star. Instead, a search on a larger area using the public RR Lyrae catalogs, we identified one known RR Lyrae who distance, metallicity, and proper motion are all consistent with Aquarius III. Therefore, this RR Lyrae, Aquarius III-V1, is the first RR Lyrae of Aquarius III despite it is located at a projected distance of $\sim 15.6r_h$ from Aquarius III. We have also ran a set of synthetic CMD, and found that UFD with properties similar to Aquarius III (and Virgo III) has a non-negligible probability to host (at least one) RR Lyrae. This finding encourages the search of RR Lyrae in other newly discovered UFD.   

\acknowledgments

We are thankful for funding from the National Science and Technology Council (NSTC, Taiwan) under the grant 113-2112-M-008-028. We sincerely thank the observing staff at the Lulin Observatory, C.-S. Lin, H.-Y. Hsiao, and W.-J. Hou, for carrying out the queue observations for this work. We thank Z.-Y. Lin and Y.-C. Pan for sharing some of the LOT time. This publication has made use of data collected at Lulin Observatory, partly supported by NSTC grant 109-2112-M-008-001. This research has made use of the SIMBAD database and the VizieR catalogue access tool, operated at CDS, Strasbourg, France. This research made use of Astropy,\footnote{\url{http://www.astropy.org}} a community-developed core Python package for Astronomy \citep{astropy2013, astropy2018, astropy2022}. 

This work has made use of data from the European Space Agency (ESA) mission {\it Gaia} (\url{https://www.cosmos.esa.int/gaia}), processed by the {\it Gaia} Data Processing and Analysis Consortium (DPAC, \url{https://www.cosmos.esa.int/web/gaia/dpac/consortium}). Funding for the DPAC has been provided by national institutions, in particular the institutions participating in the {\it Gaia} Multilateral Agreement.

The Pan-STARRS1 Surveys (PS1) and the PS1 public science archive have been made possible through contributions by the Institute for Astronomy, the University of Hawaii, the Pan-STARRS Project Office, the Max-Planck Society and its participating institutes, the Max Planck Institute for Astronomy, Heidelberg and the Max Planck Institute for Extraterrestrial Physics, Garching, The Johns Hopkins University, Durham University, the University of Edinburgh, the Queen's University Belfast, the Harvard-Smithsonian Center for Astrophysics, the Las Cumbres Observatory Global Telescope Network Incorporated, the National Central University of Taiwan, the Space Telescope Science Institute, the National Aeronautics and Space Administration under Grant No. NNX08AR22G issued through the Planetary Science Division of the NASA Science Mission Directorate, the National Science Foundation Grant No. AST-1238877, the University of Maryland, Eotvos Lorand University (ELTE), the Los Alamos National Laboratory, and the Gordon and Betty Moore Foundation.

Funding for the Sloan Digital Sky Survey IV has been provided by the Alfred P. Sloan Foundation, the U.S. Department of Energy Office of Science, and the Participating Institutions. SDSS acknowledges support and resources from the Center for High-Performance Computing at the University of Utah. The SDSS web site is \url{www.sdss4.org}.

SDSS is managed by the Astrophysical Research Consortium for the Participating Institutions of the SDSS Collaboration including the Brazilian Participation Group, the Carnegie Institution for Science, Carnegie Mellon University, Center for Astrophysics | Harvard \& Smithsonian (CfA), the Chilean Participation Group, the French Participation Group, Instituto de Astrofísica de Canarias, The Johns Hopkins University, Kavli Institute for the Physics and Mathematics of the Universe (IPMU) / University of Tokyo, the Korean Participation Group, Lawrence Berkeley National Laboratory, Leibniz Institut für Astrophysik Potsdam (AIP), Max-Planck-Institut fur Astronomie (MPIA Heidelberg), Max-Planck-Institut für Astrophysik (MPA Garching), Max-Planck-Institut für Extraterrestrische Physik (MPE), National Astronomical Observatories of China, New Mexico State University, New York University, University of Notre Dame, Observatório Nacional / MCTI, The Ohio State University, Pennsylvania State University, Shanghai Astronomical Observatory, United Kingdom Participation Group, Universidad Nacional Autónoma de México, University of Arizona, University of Colorado Boulder, University of Oxford, University of Portsmouth, University of Utah, University of Virginia, University of Washington, University of Wisconsin, Vanderbilt University, and Yale University.

Based on observations obtained with the Samuel Oschin Telescope 48-inch and the 60-inch Telescope at the Palomar Observatory as part of the Zwicky Transient Facility project. ZTF is supported by the National Science Foundation under Grants No. AST-1440341 and AST-2034437 and a collaboration including current partners Caltech, IPAC, the Oskar Klein Center at Stockholm University, the University of Maryland, University of California, Berkeley , the University of Wisconsin at Milwaukee, University of Warwick, Ruhr University, Cornell University, Northwestern University and Drexel University. Operations are conducted by COO, IPAC, and UW.

\facility{LO:1m, Gaia, PO:1.2m, PS1, Sloan}

\software{{\tt astropy} \citep{astropy2013,astropy2018,astropy2022}, {\tt dustmaps} \citep{green2018}, {\tt gatspy} \citep{vdp2015}, {\tt Matplotlib} \citep{hunter2007},  {\tt NumPy} \citep{harris2020}, {\tt PSFEx} \citep{bertin2011}, {\tt SCAMP} \citep{scamp2006}, {\tt Source-Extractor} \citep{bertin1996}, {\tt SWARP} \citep{bertin2002}}

\appendix

\section*{Identifying a Background RR Lyrae Candidate} \label{appxa}

\begin{figure*}
  \epsscale{1.1}
  \plottwo{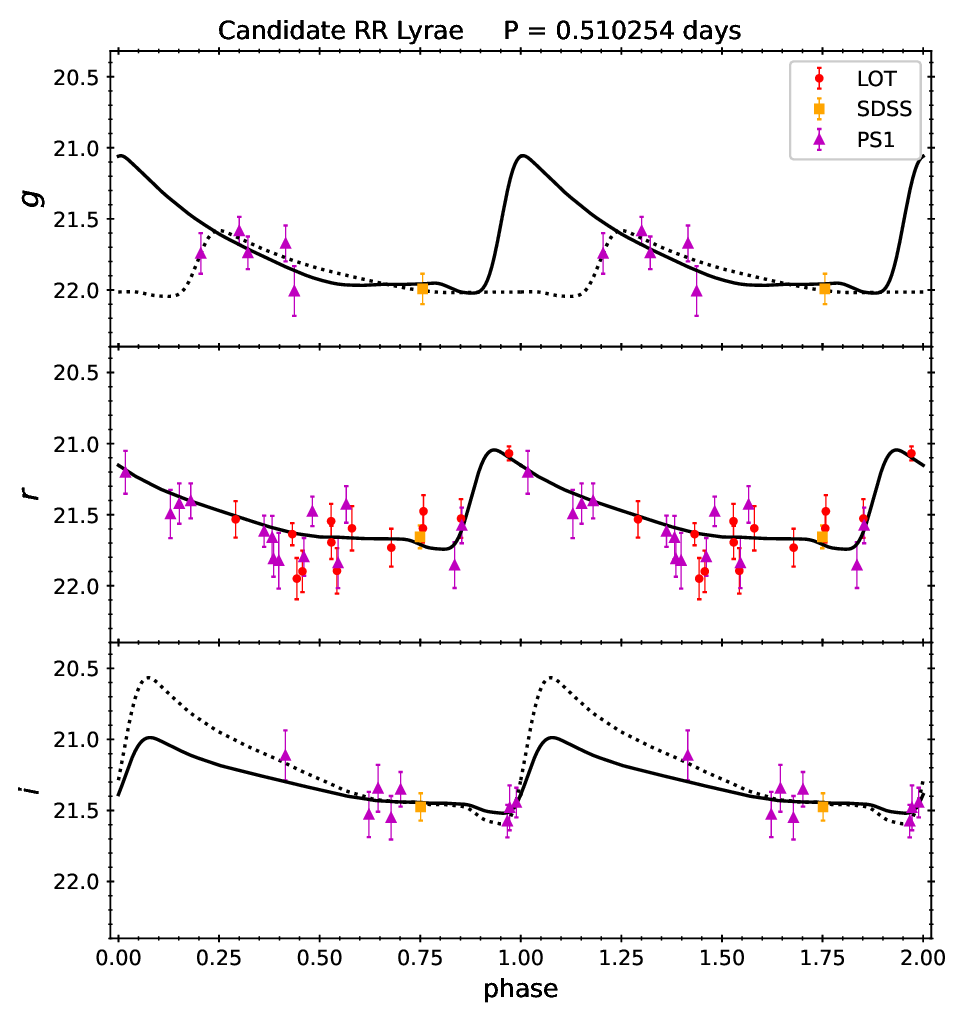}{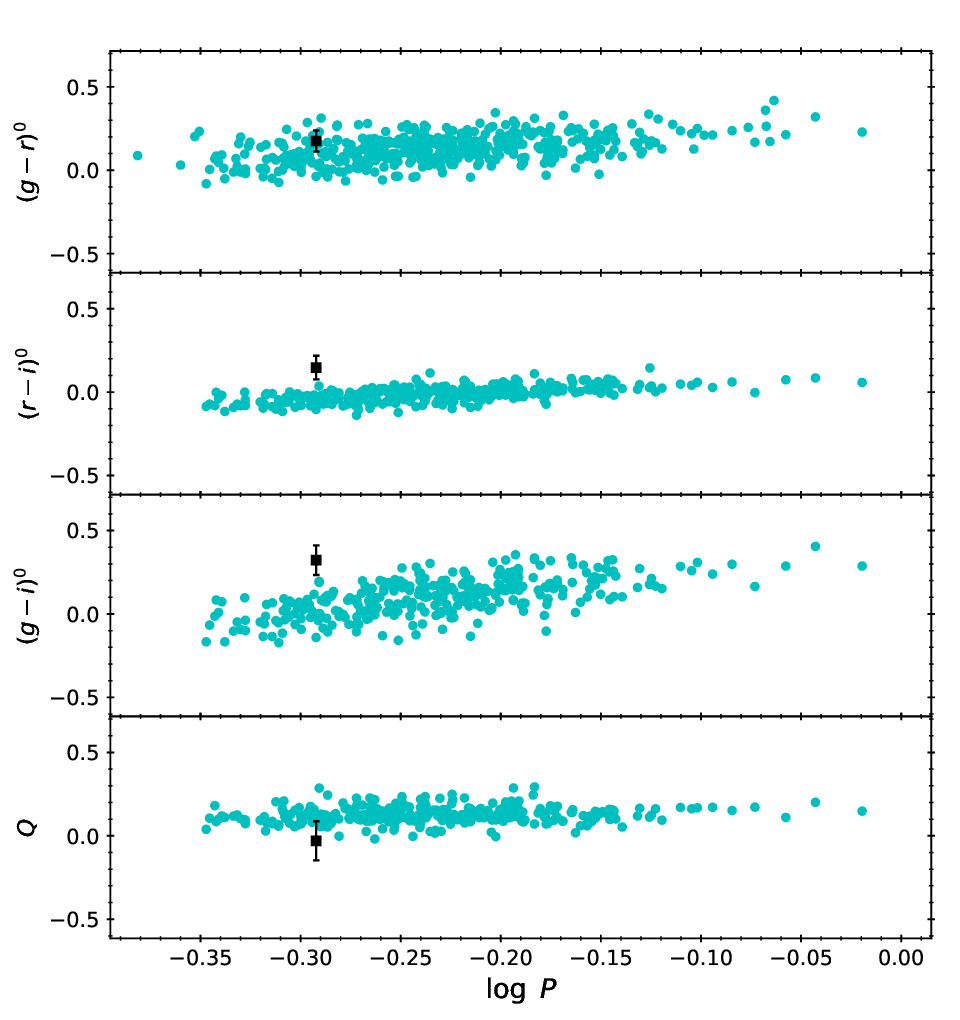}
  \caption{{\it Left Panel:} Folded light curves for the candidate variable star assuming it is an ab-type RR Lyrae. The dotted curves are the poorly fit template light curves by setting the amplitudes as a free parameter (in addition to the mean magnitude). In contrast, the solid curves are the best-fit template light curves by setting the amplitude either as a free parameter (in the $r$-band) or as a fixed parameter (in the $gr$-band; see text for more details). {\it Right Panel:} Comparison of the extinction corrected colors and the reddening-free $Q$-index for the candidate variable star to the ab-type RR Lyrae in the globular clusters \citep[cyan points; adopted from][]{ngeow2022}. The foreground extinction for this candidate variable star was obtained from using the {\tt Bayerstar2019} 3D reddening map \citep{green2019}, together with the {\tt dustmaps} \citep{green2018} code, which returned an extinction value of $0.054\pm0.004$~mag.}
  \label{fig_cand}
\end{figure*}

The SDSS DR16 objID for the candidate variable star found in Section \ref{seclot} is 1237679433443508701, which only contains 13 LOT $r$-band data points (because the brightness of this star is approaching the detection limit of our LOT observations). As a result, we supplemented the available multi-band light curves from the PS1DR2 archive with 5/14/8 data-points in the $g/r/i$-band, respectively. We further included the single-epoch $gri$-band data taken from the SDSS DR16. We have transformed the SDSS photometry to the PS1 photometric system using the transformation provided in \citet{tonry2012}, such that all photometry are in the same PS1 photometric system. We excluded the ZTFDR22 light curves, with 8/24/0 data-points, because they are noisy with large photometric errors (in the order of $\sim0.2$~mag).  

We searched the period for this candidate variable star using the multiband RR Lyrae template light curve fitting algorithm available in the {\tt gatspy} \citep{vdp2015} python package. By restricted the period range between 0.2 and 0.9~days (appropriate for RR Lyrae), we obtained a period of 0.510254~days using the $gri$-band light curves, indicating this star could be an ab-type (or fundamental mode) RR Lyrae.

The left panels of Figure \ref{fig_cand} present the folded $gri$-band light curves for this candidate variable star. We then fit the observed light curves using the RR Lyrae template light curves available from \citet{braga2024}, by setting the mean magnitudes and amplitudes as free paramters. It turned out the template light curves fit poorly on the $g$- and the $i$-band data, as indicates by the dotted curves shown in the left panels of Figure \ref{fig_cand}. In contrast, the template light curve fit the $r$-band data-points well, giving a mean magnitude of $21.496\pm0.023$~mag and an amplitude of $0.699$. Therefore, we fixed the $gi$-band amplitudes by using the $g$-to-$r$ and $r$-to-$i$ amplitude ratios as derived in \citet{ngeow2022}, and only fitting the mean magnitudes with the template light curves. The improvement of the fitted $gi$-band template light curves, displayed as solid curves in the left panels of Figure \ref{fig_cand}, can be clearly seen. The mean magnitudes in the $gi$-band were then found to be $21.720\pm0.057$~mag and $21.314\pm0.066$~mag, respectively.  

With the determined period and mean magnitudes, we compare the extinction-corrected colors of this candidate variable star to the ab-type RR Lyrae in the globular clusters \citep[adopted from][]{ngeow2022}, assuming this candidate variable star is an ab-type RR Lyrae. In addition, we have also included the reddening-free $Q$-index \citep{ngeow2022}, defined as $Q=(g-r)-1.395(r-i)$, in such comparison. Except the $(g-r)$ color, other colors and the $Q$-index for this star are marginally agree to the RR Lyrae in the globular clusters, as shown in the right panels of Figure \ref{fig_cand}. This could due to fixing the amplitudes in the $gi$-band while fitting the template light curves, and hence affecting the determination of the mean magnitudes.

Based on the extinction corrected $r$-band mean magnitude, and assumed this candidate variable star has the same metallicity as Aquarius III \citep[$\mathrm{[Fe/H]}=-2.61$~dex,][]{cerny2024}, we obtained a distance modulus of $21.06\pm0.19$~mag by using the $r$-band period-luminosity-metallicity relation derived in \citet{ngeow2022}. This distance modulus corresponding to a distance of $163.1\pm14.2$~kpc, placing this candidate RR Lyrae as a background halo star. We did not derive the distance modulus in the $gi$-band, nor using the extinction-free Wesenheit magnitudes, because of their less accurate mean magnitudes. 


\end{document}